# Simplified Graph-based Visualization for Scientific Publication

Orlando Fonseca Guilarte
Department of applied mathematics
PUC-Rio
Rio de Janeiro, Brazil
ofonsek0702@mat.puc-rio.br

Simone Diniz Junqueira Barbosa
Department of informatics
PUC-Rio
Rio de Janeiro, Brazil
simone@inf.puc-rio.br

Sinesio Pesco
Department of applied mathematics
PUC-Rio
Rio de Janeiro, Brazil
sinesio@puc-rio.br

*Abstract*—Understanding citations to scientific publications is a task of vital importance in the academic world. This task can be supported by appropriate data structures and visualization mechanisms. One challenge is the amount of existing relationships and the difficulty of determining which of the references of a document are considered the most potentially relevant to it. In this paper, we propose a simplified visualization of the relationships between scientific publications, in the form of a directed acyclic graph. From a given document, it is possible to visualize a path of references in which each step corresponds to the main citation of the previous one. A methodology is proposed in order to build this graph based in the opinion of the authors of scientific articles and an editorial board.

**Keywords—data visualization, hierarchical structure, noSQL database, interactive system**

I. INTRODUCTION

Interactive data visualizations help to clearly and efficiently interpret the underlying information, so that, through exploration, users can acquire knowledge that will assist them in making decisions. For network data, visualizations help users to perceive the relationships between items and identify the most representative ones. Graphs are used in numerous applications within the field of information visualization, enabling visual representations of the relationships between nodes in network data. They assist in the perception process and increase the level of understanding [1]. For instance, they can be used to model and visualize how similar a film is to another in terms of gender, theme, etc. [2]; to determine the structure of scientific collaborations and the status of a researcher through co-authorship networks [3] and to recommend book purchases based on the content of these and the profile of users [4].

A commonly used visualization model for academic institutions and researchers involves mapping scientific documents onto nodes of a graph, and citations or bibliographic references onto relationships between those nodes [5, 6, 7, 8]. This model is useful because the set of citations is an indicator of the quality of scientific articles, by both the quantity and the selection of citations, which can be analyzed in the graph. The authors of scientific documents are usually careful in selecting the works that will be listed. This is crucial because in only a small sample they must provide which articles are potentially relevant to their research to cite them in the development of the document, as seen in [9]. The list of references is an index of the selectivity and moderation where authors have filtered the trivial information from the relevant one according to [10], but in several occasions the amount of irrelevant references provided by authors is excessive.

For most papers, there are several reasons to cite a research paper. For example, the author could list some scientific documents because they support or criticize the idea presented in it, as seen in [11]. The user may also have other reasons to cite, which might even lead to irrelevant citations. For instance, the most recognized articles in the area may be cited with the aim of gaining credibility or because the author has spent some time studying and understanding them, even if they are not directly related to the current paper. Self-promotion in citations can also occur, which means that the authors list their previous publications to gain visibility.

In this paper we consider that only up to three bibliographical references are really motivated and have contributed the main ideas for a paper. Those references can be categorized as PBas, PModi or PUse. In PBas, the author uses the cited work as a starting point; in PModi they have adapted or modified algorithms or tools; and in PUse they have used algorithms and data from previous works. The remaining citations can be categorize as neutral (Neut) or irrelevant (Weak). These categories are explained in [12].

Representing and viewing papers and their citation relationships as graphs can be confusing, because there are usually many citations to each node of the graph. Therefore, it can be arduous to find a sequence of closely related works, i.e., works with relevant citation relationships between them. This is like that because only the author or experts in the subject can truly affirm the reasons why the references were included and, therefore, decide which of them should be considered the most important to visualize. Although there have been proposed algorithms to analyze the classification or ranking of references, as in [7, 13], we considere that the opinion of the author or of an expert is essential in this decision.

For the reasons mentioned above, we have developed a web application that presents a simplified visualization of the relationships between scientific publications, in the form of a directed acyclic graph.

The main contribution of this work lies in considering to reduce the number of edges of the graph represented by scientific documents, as a result of assessing the relevant citations. Even though not all the references are shown, the user can obtain a less cluttered visualization of the most relevant relationships between them.

We also propose a methodology to build such a structure, based on the suggestion of the authors and the critical evaluation of experts, assigning credibility to the visual result.

Because of the chronological nature of the data, we have arranged the graph nodes in levels, each one corresponding to a time interval (5 years in the example of figure 5). We have also used visual variables in each node to emphasize technical information of the article, such as number of authors, number of citations, etc.

This paper is structured in four sections, starting with those works directly related to the research topic and then with the explanation of the chosen data sets. Then we present our proposed solution and finally the results of a user study. The conclusions and future work complete this paper.

## II. RELATED WORKS

In recent years, several papers have presented a study of scientific publications, their citations and graphic representations in the form of graphs.

Sibaroni et al. [13] propose to build relationships between documents using basic approaches like analysis of co-citations, reference lists and bibliographic coupling. Singh et al. [14] point out that the quality of the citations is very important, and that it is not enough to consider only their quantity, as it has been done with metrics like h-index and impact factor.

Several visualizations of scientific documents have been proposed to help researchers explore graphs of references. PyScholarGraph [15] shows a framework for indexing, searching and viewing references, based on data recovered from the CiteSeerX repository and indexed in a graph-oriented database. Waumans and Bersini [6] adopt an evolutionary perspective, building the graph in the form of a genealogical tree, showing the rate of growth of the number of citations for each article. In PyGraphviz [5], a graph was created to represent the evolutionary process of the deep learning field in the last 25 years. Wei et al. [7] propose an interactive visualization method that uses citation paths, as a result of a variation in the TF-IDF scheme. As shown in [16], the data can be processed previously and sent to a database with the information of scientific journals, authors and research documents, from where they can be consulted using the Cypher language [17].

However, in [16] a hierarchical structure to facilitate user understanding is not presented. In [6], although such a structure is considered, the analysis period is not segmented into time intervals. In [15, 7, 14] the opinion of specialists in the area was not considered when building the graph. Finally, in [5] it is difficult to visually obtain a sequence of closely related works due to the number of relationships that it presents.

Until now, we do not find in the literature an interactive visualization of scientific publications that shows data in a simple and comprehensive way for users, with defined years and a hierarchical structure of potentially relevant documents without requiring calculations of similarity among them but including the opinion of experts. In our proposed representation, from any publication in the graph to another (typically older) publication, there will be only one edge, to help users understand the main source of the publication, e.g., the origin of the addressed problem. In the same way a scientific document can be considered as the root of the tree because it is old but important, there can be several sequences that show the evolution of the research field to this day. For this purpose, the present research was developed.

## III. DATA SET

The data collections we chose to visualize with our proposed approach are the following: vispubdata [18] and vismc.

The vispubdata set contains information about the publications in IEEE Visualization (IEEE VIS). Each scientific document in our system will be identified by its DOI and will contain data such as the names of the authors, year of publication at the conference, title, abstract, as well as a list of citations to other documents of the conference. This study will include papers published in recent years and that have been relevant to the community as they were cited by others.

The vismc set contains a particular selection of articles published in various sites that address the issue of Marching Cubes, computer graphic topic on the extraction of isosurfaces, explained in [19]. The data is in .csv format and it is partially illustrated in Table I of the appendix.

## IV. SIMPLIFIED GRAPH-BASED VISUALIZATIONS

### A. Description

The proposal presented in this paper is based on the following four characteristics:

1) The data structure to represent the network of citations is a directed acyclic graph, where the nodes are the scientific documents and the edges are the citation relations between them.
2) Each node related to an article appears only once in the representation.
3) The graph is organized in levels defined according to the date of publication of the nodes.
4) Each node has at least one edge emerging from it.

We decided to develop a directed graph because the citation has a source node and destination node that are not interchangeable. We avoid cycles by guaranteeing that a document can only refer to another document that already existed.

A document is not represented multiple times in the graph because it would increase clutter and make visualization difficult. Even if the graph has connected components of very different topics, a node will be connected to only one other node, which will be the one that best characterizes it.

A natural way to observe the progress of different lines of research is to have a hierarchical structure, as observed in [6]. To reduce clutter in the tree, we define levels in the graph with time intervals: the oldest articles will be located at the top of the graph and the most recent ones at the bottom.

If it is considered that basically each node will have only one edge emerging from it, then the resulting structure will be a directed tree. It is a variant that offers visual simplicity and better exploration to the data.

Having only one edge emerging from the node, despite being a simplification, does not omit the most important information because it determines the most important scientific article that inspired that research. It is possible to perceive this information in a visual way more quickly and easily, without losing the essence of finding good citations from a specific publication.

We argue that the exploration of the tree to find a document is more convenient. The user will not only find documents on the same subject, but he/she will have the most relevant documents, in sequential order of citations. A document can be found by searching its DOI, or by visually inspecting the most cited nodes in the area. It is also possible to search a document by its author's name or by its title. One can see they may be on one or more paths or branches of study. With this proposal the user would obtain a visual recommendation of which scientific documents should be studied or considered for inclusion in the list of references in their future papers.

In the visualization, the nodes are circles whose color and position reflect the year of publication, so that those published in a defined interval of years will be on the same level and with the same color. The circle size indicates the in degree (the number of incident edges) of the corresponding document, and its thickness indicates the number of authors. Each edge, referring to a citation relationship, is represented as a directed line from a more recent node to an older one. Figure 1 shows those visual features. We can see that the document called "Discretized marching cubes" has the greatest number of authors within those presented and the document called "Marching Cubes" is the one that has more citations to it, and it is the oldest document in the structure.

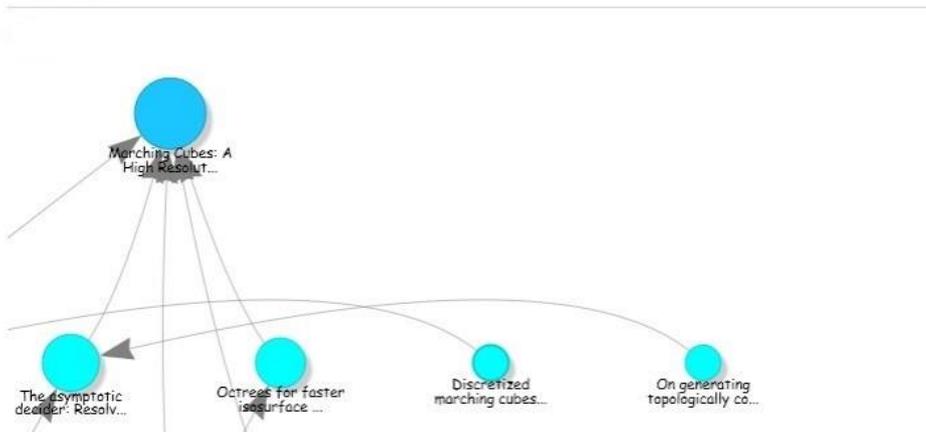

Fig. 1. Example of the visualization.

## B. Implementation

All the information collected from the different data sets are stored as a graph in Neo4j [17], a NoSQL database designed for graphs. Through the declarative language Cypher, the database is queried, allowing the manipulation and analysis of data in an easy and intuitive way.

We chose the Python programming language and its Django framework [20] for developing the web application. The selection was justified because Python has very good libraries to interact with Neo4j, like py2neo. Django was chosen because of its work philosophy. It is important to note that it was also possible to interact with the database from the templates and views of Django, with the official Neo4j driver called neo4j-driver. For the interactive visualization, the JavaScript library named vis.js was used [21]

To create the graph, the following procedure is presented:

1) For each one of the existing documents in the database:
   a) The visual features of the document are established, according to the corresponding information.
   b) The referenced document is identified to establish the visual features of the relationship.
2) A hierarchical architecture of the graph is established, as well as events, physical features and animation.

## V. EMPIRICAL STUDY

At first, the resulting graph is shown when the visualization procedure is used for the vispubdata data set. The selection of the most representative reference for each document was determined randomly among the document references that were published in the same conference.

Figure 2 shows a section of the graph that displays some of the most relevant scientific documents of recent years in the area, totaling 596 nodes and 524 relationships.

For comparative purposes in terms of the number of edges, Figure 3 shows how difficult it is to visualize the data and the exploration of the graph, although it has a hierarchical structure and organized in levels by publication year. In total there are 596 nodes and their respective reference lists published in the conference, totaling 1243 relationships.

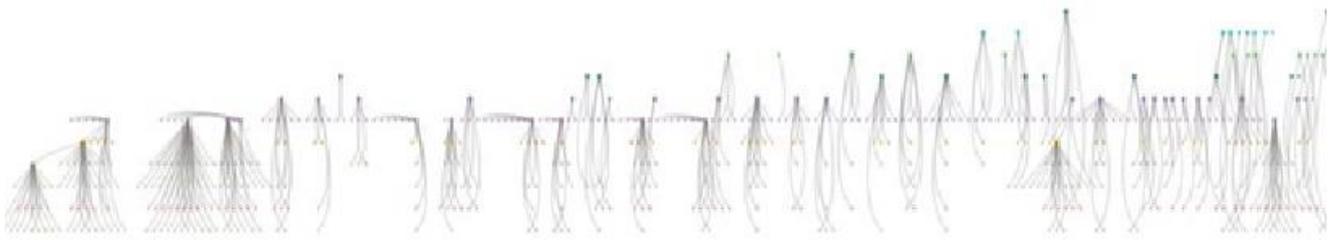

Fig. 2. Simplified hierarchical structure by considering a relationship as being the most relevant for each node.

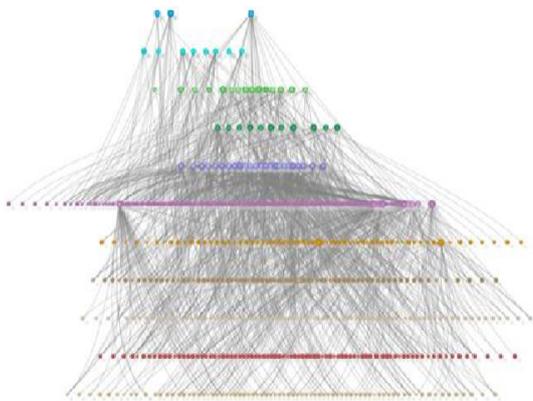

Fig. 3. Hierarchical structure considering several citation relationships for each node.

The procedure to build the graph of scientific documents according to the proposal presented in this paper is specified below:

1) The author (registered user) submits the data of a new scientific document in to the system, to be included in the graph. He/She also includes 1-3 ordered suggestions of references that, in his/her opinion, are the most relevant ones for the paper. The user cannot be the author of all references.

2) An expert (a designated editor-in-chief) names a few researchers to analyze the references recorded in the previous step. In case those researchers disagree with the suggestion given by the author, they could choose another as the most relevant one, according to their experience in the area.

3) An interactive node is created with the metadata of the scientific article.

4) The relationship between the node and its main reference is created, according to the decision of the experts (based on the authors' suggestion).

5) The data corresponding to the step 3 and 4 are stored in the graph database.

The vismc data set was chosen to evaluate the building process of the graph, following the criteria of an expert in the area. In general, the graph contains 17 nodes and multiple relationships to the other nodes in the set, totaling 43 relationships. Figure 4 shows the structure of the graph with its citation relationships.

Even with a limited data set, the task of finding a sequence of potentially relevant documents from, for example, a node at the bottom-most level, can be difficult to determine visually.

Figure 5 shows the same 17 nodes of the set with their visual features, but now each node has only one relationship, deemed the most relevant one by an expert. In total, there are 16 relationships between the nodes.

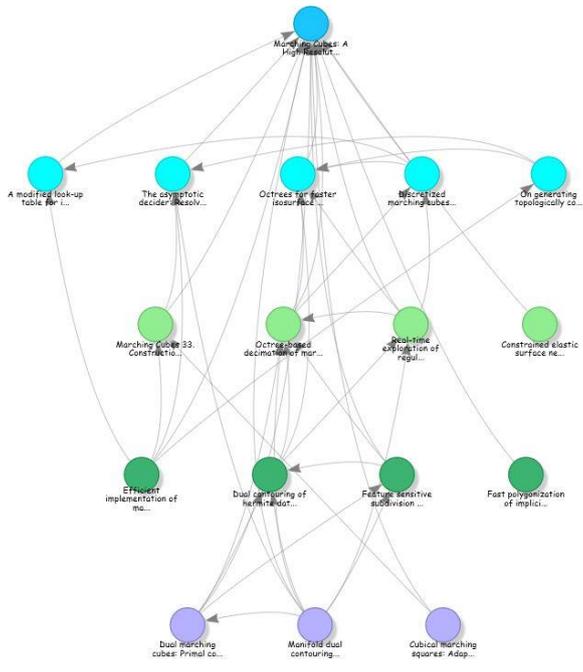

Fig. 4. Hierarchical structure with all relationships.

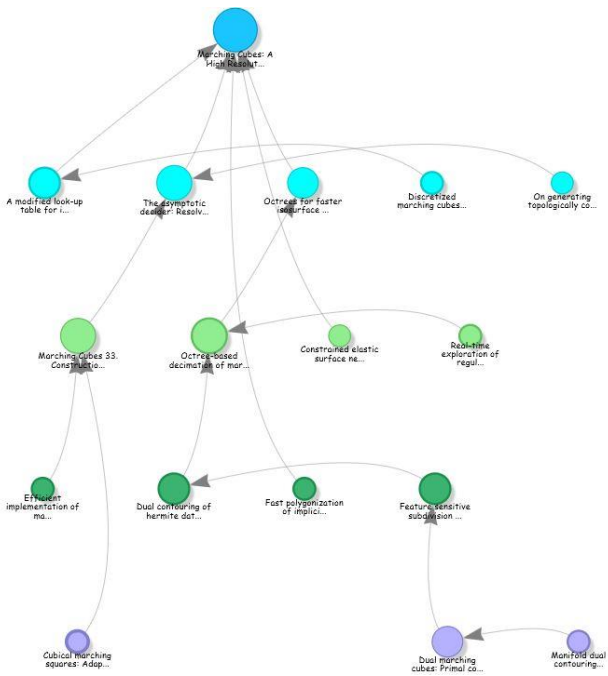

Fig. 5. Hierarchical structure with the participation of an expert to determine the potentially most relevant relationship for each node.

To evaluate the proposed visualization methodology, we conducted an empirical study in an academic environment, where the participants could indicate how much the visualization presented in this paper was useful and easy to interpret (or not). We expected that obtaining a set of relevant documents to study is easier with a simplified visualization than with another that shows all the relationships. These studies are a widely used form of evaluation by offering a scientifically sound method to measure the performance of visualizations, as explained in [22].

The study participants were 13 people from the academic and scientific fields: 10 graduate students and 3 researchers. The participation was voluntary and each participant signed an informed consent form. After using the system, each participant answered a questionnaire with 9 questions. For the first question, a multiple-choice question with a single answer was used, and for the remaining questions, a 7-point Likert scale was used, which varied from Completely Disagree to Completely Agree. The questionnaire was designed to assess not only to what extent the visual application was useful, but which of the visualizations were considered the most beneficial and convenient.

First, the interviewer presented a scenario to motivate the participant to think of the minimal set of the most relevant bibliographical references of their last work. Then, he started with question 1 (all questions are shown below). Then he presented the web application, with two data visualizations for comparative purposes: visualization A presents a hierarchical structure with several relationships from each node; and visualization B presents a hierarchical structure with only one relationship from each node, representing the most relevant citation it makes. The interviewer also explained the visual representation of each scientific document and its relationships. He proposed the task of exploring the two visualizations and look for any document to obtain their information. The participant would then answer questions 2 and 3, regarding the ease of use of the tool. Finally, the participant answered questions 4-9, related to both visualizations, and without any explanation or help from the interviewer.

The questionnaire comprised the following questions and statements:

1) Imagine that a new requirement of the peer-reviewed journals to accept articles is a significant reduction of the bibliographic references. You need to select the most relevant references, which are more closely related to the research. What is the minimum acceptable number, according to your most recent article?

2) I understand very well how the tool is to be used.

3) I understand very well when and why to use the tool.

4) I can easily determine visually any document cited by another document.

5) I can easily determine visually a sequence of documents in which each document corresponds to the main citation of the previous one.

6) It would be easy to identify the main bibliographical references of any document in the bottom-most levels.

As we needed to ask questions 4, 5 and 6 about both visualizations, they were asked as questions 4-6 for visualization A, and duplicated as questions 7-9 for visualization B.

The results of the questionnaire are represented in two graphs. Figure 6 shows the graph for question 1, where the first bar represents the number of participants who answered that it is possible to reduce to up to 3 the number of potentially relevant references, and the second bar represents those who answered that this amount should be greater than 3. Figure 7 shows a stacked bar graph for questions 2 to 9, where each bar segment represents the number of participants who gave the same answer, and the colors of the bars represent the options according to the defined scale. This varies from Completely Disagree (dark red color) to Completely Agree (dark blue color).

Most participants (including all the researchers) could reduce the number of most relevant references to fewer than 3. The tool was considered easy to use, and more than 84 % of the participants stated they would know when and why to use it. For visualization A (questions 4, 5 and 6), the number of participants who answered Completely Agree was on average 1.33, a much lower number than the average of those who answered Completely Agree for visualization B (questions 7, 8 and 9), which was of 11.66. The study results provide evidence that the participants liked the proposed methodology, and therefore it is worth advancing the research.

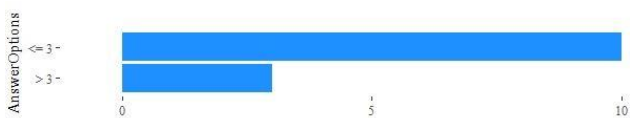

Fig. 6. Answer to the question 1.

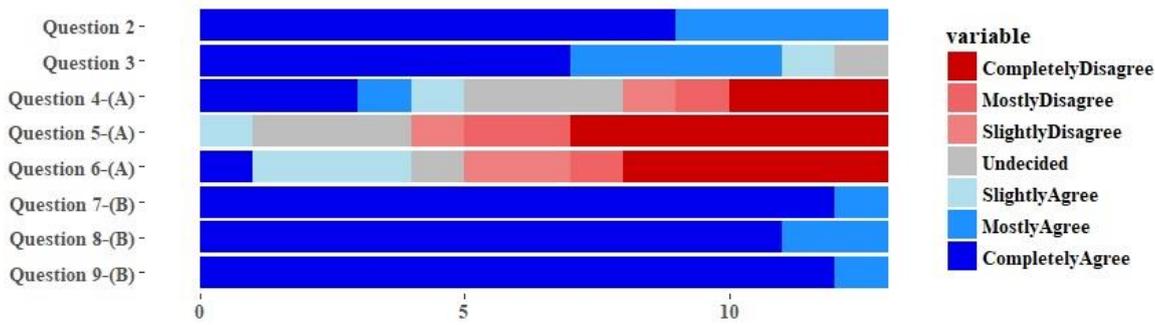

Fig. 7. Answer to questions 2 to 9.

## VI. CONCLUSIONS AND FUTURE WORK

In this research, a new methodology was proposed to visualize the relationships between scientific publications, in the form of a directed acyclic graph. Through the evaluation of experts, it was possible to determine the most relevant reference of a document. This led to have a simplified model that helps users to visually explore the graph, making it possible and easy to find sequences of closely related research. Two sets of data were visualized using comparative approaches (having several citations for each document versus having only the most relevant citation for each article). An empirical study was conducted to evaluate the proposed methodology.

As future work, we intend to evaluate the procedures with other data sets and compare the results with other methods to select potentially relevant references for each document by using algorithms for this purpose, such as TF-IDF or LDA. Other algorithms to determine the number of levels of the graph for years can be implemented, depending on the number of publications in the database for those years. This research could also complement the work of editors and reviewers of scientific journals.

## APPENDIX

TABLE I. SOME DATA OF THE VISMC SET

| id | title | DOI | authors | year | abstract | keywords | url | ref |
|---|---|---|---|---|---|---|---|---|
| 1 | Marching Cubes: A High Resolution 3D Surface Construction Algorithm | 10.1145/37402.37422 | W. E. Lorensen; H. E. Cline | 1987 | We present a new algorithm, called marching cubes, that creates triangle models of constant density surfaces from 3D medical data… | computer graphics, medical imaging, surface reconstruction | http://marchingcubes.org/images/f/f9/MarchingCubes.pdf | |
| 2 | A modified look-up table for implicit disambiguation of Marching Cubes | 10.1007/BF01900830 | C. Montani; R. Scateni; R. Scopigno | 1994 | A new triangulation scheme for the Marching Cubes algorithm is proposed… | volume visualization, surface reconstruction, look-up tables, marching cubes algorithms | https://link.springer.com/article/10.1007/BF01900830 | 1 |

| 3 | The asymptotic decider: Resolving the ambiguity in marching cubes | 10.1109/VISUAL.1991.175782 | G. M. Nielson; B. Hamann | 1991 | A method for computing isovalue or contour surfaces of a trivariate function is discussed… | | https://graphics.ethz.ch/teaching/scivis_common/Literature/TheAsymptoticDecider.pdf | 1 |
| 4 | Octrees for faster isosurface generation | 10.1145/78964.78965 | J. Wilhelms; A. Van Gelder | 1992 | The large size of many volume datasets often prevents visualization algorithms from providing interactive rendering… | languages, theory, negation as failure, well-founded models, xpoints, unfounded sets, stable models | https://users.soe.ucsc.edu/~avg/Papers/oct92.pdf | 1 |